\documentclass[floatfix,aps,twocolumn,amsmath,amssymb]{revtex4}
\pdfoutput=1

\usepackage{amsfonts, amssymb, amsmath, dsfont}
\usepackage{graphicx}
\usepackage{float}
\usepackage{color}
\usepackage{ulem}
\usepackage[utf8]{inputenc}
\usepackage[T1]{fontenc}
\usepackage{hyperref}
\usepackage[svgnames]{xcolor}
\hypersetup{colorlinks,urlcolor=DarkBlue,linkcolor=DarkBlue,citecolor=DarkBlue,pdfdisplaydoctitle=true,pdfpagemode=UseOutlines,bookmarksnumbered=true,bookmarksopen=true} 

\newcommand{\bm}[1]{\boldsymbol{\mathbf{#1}}}

\newcommand{\Ee}{\operatorname{E}}
\newcommand{\Ve}{\operatorname{Var}}

\newcommand{\de}{\operatorname{d}}

\newcommand{\eq}[1]{Eq.~\eqref{#1}}

\newcommand{\Eq}[1]{Equation~\eqref{#1}}

\newcommand{\fig}[1]{Fig.~\ref{#1}}

\newcommand{\Fig}[1]{Figure~\ref{#1}}
\newcommand{\tab}[1]{Table~\ref{#1}}

\newcommand{\beginmain}{%
		\renewcommand{\thesection}{\Roman{section}}
     }

\begin{document}

\title{Cram\'er-Rao analysis of lifetime estimations in time-resolved fluorescence microscopy}

\author{D. Bouchet}
\author{V. Krachmalnicoff}
\altaffiliation{valentina.krachmalnicoff@espci.fr}
\author{I. Izeddin}
\altaffiliation{ignacio.izeddin@espci.fr}

\affiliation{Institut Langevin, ESPCI Paris, CNRS, PSL University, 1 rue Jussieu, 75005 Paris, France}

\begin{abstract}
Measuring the lifetime of fluorescent emitters by time-correlated single photon counting (TCSPC) is a routine procedure in many research areas spanning from nanophotonics to biology. The precision of such measurement depends on the number of detected photons but also on the various sources of noise arising from the measurement process. Using Fisher information theory, we calculate the lower bound on the precision of lifetime estimations for mono-exponential and bi-exponential distributions. We analyse the dependence of the lifetime estimation precision on experimentally relevant parameters, including the contribution of a non-uniform background noise and the instrument response function (IRF) of the setup. We also provide an open-source code to determine the lower bound on the estimation precision for any experimental conditions. Two practical examples illustrate how this tool can be used to reach optimal precision in time-resolved fluorescence microscopy.
\end{abstract}

\beginmain

\maketitle

\section{Introduction}

Nowadays, the precise determination of the lifetime of fluorescent emitters has become essential for a wide range of applications. Indeed, enhancing the spontaneous rate of single emitters is a topical challenge in nanophotonics \cite{novotny_antennas_2011,koenderink_single-photon_2017}, leading to the measurement of strongly reduced fluorescence lifetimes \cite{Kinkhabwala_large_2009,belacel_controlling_2013,akselrod_probing_2014,hoang_ultrafast_2016,bidault_picosecond_2016,cambiasso_bridging_2017}. In biology, the contrast induced by lifetime variations is
used to map different parameters on biological samples \cite{gadella_fluorescence_1993,berezin_fluorescence_2010,becker_fluorescence_2012} such as the viscosity, the potential of hydrogen (pH) or the interaction between two emitters due to F\"orster resonance energy transfer (FRET). Whatever application one is interested in, the best precision that can be achieved depends on both the experimental conditions and the efficiency of the estimators. While the performances of lifetime estimators can numerically be studied using Monte Carlo experiments \cite{bajzer_maximum_1991,tellinghuisen_bias_1993,sharman_error_1999,rowley_bayesian_2011,kaye_developing_2017}, finding the right experimental conditions requires a reliable benchmark to compare the performances of different experimental setups. Such a benchmark is set by Fisher information theory and the Cram\'er-Rao inequality, which gives the lower bound on the variance of unbiased estimators \cite{kay_cramer-rao_1993}. In other words, the Cram\'er-Rao inequality allows one to calculate the best precision that can be achieved in the estimation of one or several parameters, taking into account the various constraints induced by an experiment. In the recent years, this gauge became a standard used to assess the limit of localisation precision in the context of single-molecule microscopy \cite{ober_localization_2004,mortensen_optimized_2010,deschout_precisely_2014,chao_fisher_2016,bouchet_probing_2019}. Other recent applications of this formalism include the investigation of the dynamics of single molecules on the millisecond time scale \cite{watkins_information_2004} and the comparison of different imaging modalities in fluorescence diffuse optical tomography \cite{boffety_analysis_2008,boffety_cramer-rao_2011}. Among the different techniques that can be implemented for lifetime measurements \cite{gratton_fluorescence_2003,suhling_time-resolved_2005,becker_fluorescence_2012}, TCSPC is commonly used to exploit low level light signals with picosecond resolution \cite{becker_advanced_2015}. Using the Cram\'er-Rao inequality, a relation between the estimation precision and the number of collected photons was obtained in 1992 for a simplified TCSPC model \cite{kollner_how_1992}. Here, we perform an extensive Cram\'er-Rao analysis to unravel the dependence of the lifetime estimation precision on experimentally relevant parameters, in the general case of a bi-exponential distribution along with any non-uniform background signal and considering the finite IRF of the setup. In addition, we provide an open-source code (Code File 1, see \cite{code_ref}) which computes the Cram\'er-Rao bound for any set of experimental parameters. Within this framework, we thus provide a versatile tool which can be used for different purposes such as determining the shortest lifetime that can be probed with a TCSPC setup or achieving optimal contrast for FLIM-based applications.

\section{Results and discussion}

\subsection{Excited-state lifetime distribution}

Let us consider that the measured decay histogram follows a bi-exponential distribution with an additional contribution due to background noise. Usually this noise originates from dark counts due to the detection process, unfiltered excitation laser or luminescence of the substrate. Assuming that the noise follows a known probability density function (PDF) noted $q_b(t)$, the total signal can be modeled by using a set of 5 unknown parameters noted $\bm{\theta}$, namely, the decay rate of each component ($\Gamma_1$ and $\Gamma_2$), the average number of detections for each component ($N_1$ and $N_2$), and the average number of detections due to background noise ($N_b$). Furthermore, we consider that the excitation laser has a finite pulse duration, and that the jitter of the detection system induces a loss of precision over the photon detection time. These two effects can be accounted for by measuring the IRF of the system, which is described by a PDF noted $q_{irf}(t)$. Then, if the detected photon rate does not exceed the maximum counting speed of the detector, the expected number of events $f_i$ detected in the $i$-th bin of the decay histogram (and associated with delays in-between $t_i$ and $t_{i+1}$) reads
\begin{equation}
\begin{split}
f_i =& N_1 \sum_{l=0}^{+ \infty} \int\limits_{t_i+lT}^{t_{i+1}+lT} \left[ q_{irf}(t) \ast \Gamma_1 e^{-\Gamma_1 t} \right] \de t \\
&+ N_2 \sum_{l=0}^{+ \infty} \int\limits_{t_i+lT}^{t_{i+1}+lT} \left[ q_{irf}(t) \ast \Gamma_2 e^{-\Gamma_2 t} \right] \de t \\
&+N_b \int\limits_{t_i}^{t_{i+1}} q_b(t) \de t \; ,
\end{split}
\label{eq2}
\end{equation}
where $T$ is the repetition period of the excitation laser. In this equation, only the first term of the sums (corresponding to $l=0$) is significant if the fluorescence lifetimes associated with both exponential decays are shorter than the repetition period. For a given integration time, and assuming that the detections are independent, we can then model the distribution of detected events (including fluorescence photons and background noise) for each point of the decay histogram by a Poisson distribution of expectation $f_i$. The PDF associated with the observation of $X_i$ events on a given point of the decay histogram is therefore expressed by
\begin{equation}
p_i(X_i; \bm{\theta}) = \frac{f_i^{X_i}}{X_i!}e^{-f_i} \; .
\label{eq3}
\end{equation}

\subsection{Cram\'er-Rao lower bound}

In estimation theory, a well-known result is that the variance of any estimator $\bm{\hat{\theta}}$ must satisfy the Cram\'er-Rao inequality \cite{kay_cramer-rao_1993}, which reads 
\begin{equation}
\Ve (\hat{\theta}_j) \geq \left[\bm{\mathcal{I}}^{-1}(\bm{\theta})\right]_{jj} \; ,
\label{eq4}
\end{equation}
where $\Ve$ is the variance operator and $\bm{\mathcal{I}}$ is the Fisher information matrix defined by 
\begin{equation}
\left[\bm{\mathcal{I}}(\bm{\theta})\right]_{jk} = \Ee \left[ \left( \frac{\partial \ln p(\bm{X} ; \bm{\theta})}{\partial \theta_j} \right) \left( \frac{\partial \ln p(\bm{X} ; \bm{\theta})}{\partial \theta_k} \right) \right] \; ,
\label{eq5}
\end{equation}
where $\Ee$ is the expectation operator. The Fisher information matrix can be interpreted as a measure of the amount of information about the parameters $\bm{\theta}$ contained in a given data set $\bm{X}$: the more information about the parameters, the lower the bound on the variance of their estimators. In the case of TCSPC measurements, we can assume that the $n$ data points of the decay histogram are independent. Moreover, since $p_i(X_i; \bm{\theta})$ is a Poisson distribution of expectation $f_i$, the variance of this distribution is also equal to $f_i$. From \eq{eq5}, we obtain
\begin{equation}
\left[\bm{\mathcal{I}}(\bm{\theta})\right]_{jk} = \sum_{i=1}^n \frac{1}{f_i}\left( \frac{\partial f_i}{\partial \theta_j} \right) \left( \frac{\partial f_i}{\partial \theta_k} \right) \; .
\label{eq7}
\end{equation}

The magnitude of the off-diagonal elements determines the extent to which the Cram\'er-Rao lower bound on a given parameter is affected by the estimation of the other parameters. While these off-diagonal elements generally vanish in the context of single-molecule localisation \cite{deschout_precisely_2014,chao_fisher_2016}, the cross-terms of the information matrix must be considered here. Indeed, the precision of lifetime estimations can be strongly influenced by a lack of information about other parameters. This is notably the case for bi-exponential decay histograms characterized by two decay rates of the same order of magnitude ($\Gamma_1 \sim \Gamma_2$).

Dimensionless quantities for the parameters involved in the calculations can be obtained by performing the change of variable $u =\Gamma_1 t$, and by normalizing parameters by $N_1$ and $\Gamma_1$ (this choice was notably made in \cite{kollner_how_1992}). Hence, we define the normalized repetition period $r=\Gamma_1 T$, the number of data points per period $k=n/r$ and the normalized expected number of counts due to background noise $\beta=N_b/(rN_1)$. As an example, let us consider a common organic dye, Alexa Fluor 488 ($\tau = 4.1$~ns) from which $2,000$ emitted photons have been collected by the detection system. Assuming a repetition rate of 80~MHz, a board resolution of 16~ps and $1,000$ detections due to background noise, the value taken by the dimensionless parameters $r$, $k$ and $\beta$ are respectively of the order of $r = 3$, $k = 256$ and $\beta = 0.16$. Bi-exponential decays can be parametrized using the ratio of the decay rates $\gamma=\Gamma_2/\Gamma_1$ and the ratio of the expected number of detections $\eta=N_2/N_1$. \tab{table1} summarizes the parameters used in the model. With these parameters, \eq{eq4} reads
\begin{equation}
\frac{\sigma_{\Gamma_1}}{\Gamma_1} \geq \frac{1}{\sqrt{N_1}} \times F \left(\eta,\gamma,r,k,\tilde{q}_{irf},\beta,\tilde{q}_{b} \right) \; ,
\label{eq9}
\end{equation}
where $\sigma_{\Gamma_1}$ is the standard error on the decay rate estimates and $F$ can be calculated by numerical inversion of the information matrix, the elements of which are reported in Section~1 of \textit{Appendix: Numerical methods}. It must be noted that the Cram\'er-Rao bounds are the same for the relative standard error on the decay rate and lifetime estimators (see Section~2 of \textit{Appendix: Numerical methods}).

\begin{table*}[htpb]
\begin{center}
\caption{Parameters Involved in the TCSPC Data Model.}
\medskip
\begin{tabular}{|c|c|c|}
 \hline
 & Parameters & Dimensionless parameters \\
 \hline
First fluorescence decay & $N_1$ and $\Gamma_1$ & \\
Second fluorescence decay & $N_2$ and $\Gamma_2$ & $\eta=N_2/N_1$ and $\gamma=\Gamma_2/\Gamma_1$ \\
Repetition period & $T$ & $r=\Gamma_1 T$ \\
Number of data points & $n$ & $k=n/r$ \\
Instrument response function & $q_{irf}(t)$ & $\tilde{q}_{irf}(u)= q_{irf}(u/\Gamma_1) / \Gamma_1$ \\
Background noise & $N_b$ and $q_b(t)$ & $\beta=N_b/(rN_1)$ and $\tilde{q}_b(u)=q_{b}(u/\Gamma_1) / \Gamma_1$ \\ 
 \hline
\end{tabular}
\label{table1}
\end{center}
\end{table*}

\Eq{eq9} explicitly gives the fundamental limit on the precision of decay rate (and lifetime) estimations. In these expressions, $F$ is calculated by inverting the information matrix and describes the influence of the different parameters involved in the model on the value of the Cram\'er-Rao bound. A low value of $F$ indicates an experimental setup with a high sensitivity: the $F$-value is always greater than unity and equals unity when the shot noise limit is reached. Optimisation of a TCSPC setup is therefore achieved when the $F$-value reaches unity, indicating that the precision of lifetime estimation is limited by the number of detected fluorescence photons. For these reasons, the $F$-value is used as a figure of merit to quantify the performance of a lifetime imaging technique \cite{gerritsen_fluorescence_2006,philip_theoretical_2003}. In the following sections, we will perform a parametric study of the $F$-value. To do so, we will consider as a reference situation the ideal case for which $k=500$, $r=100$, $\beta=0$ and $\tilde{q}_{irf}$ is a Dirac delta function. With these parameters, the $F$-value is approximately of unity. Each parameter will then be individually varied, in order to highlight the influence of each parameter upon the $F$-value.

\subsection{Mono-exponential case}

\begin{figure}[htbp]
\begin{center}
\includegraphics[width=\linewidth]{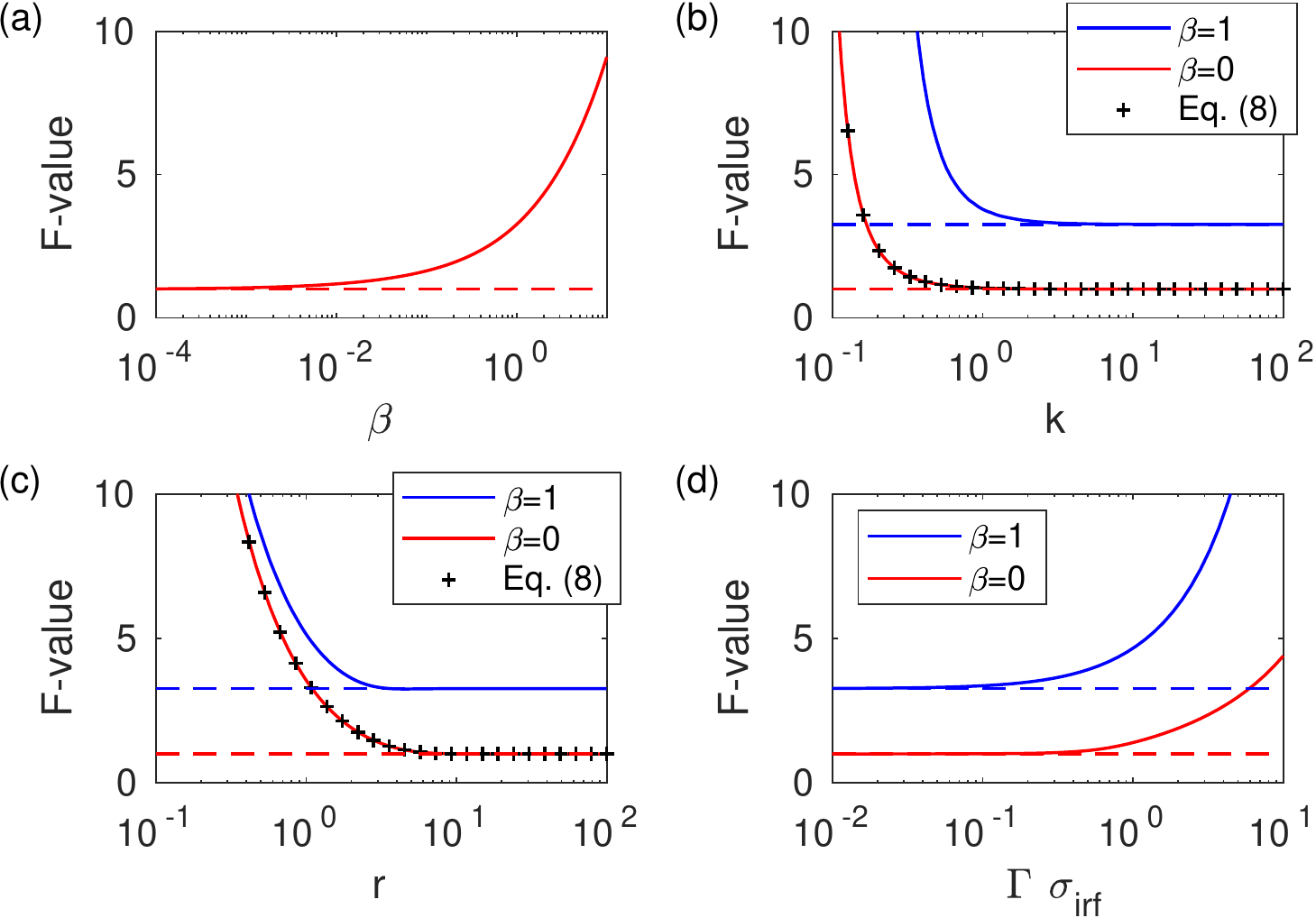}
\end{center}
\caption{Precision of lifetime estimations as a function of (a) the number of counts due to background noise, (b) the number of data point, (c) the repetition period and (d) the standard
deviation of the IRF. Each parameter is individually varied with respect to the ideal case in which $k = 500$, $r = 100$, $\beta = 0$ and $\tilde{q}_{irf}$ is a Dirac delta function. Dashed lines represent the asymptotic values.
}
\label{fig1}
\end{figure}

Let us consider a signal following a mono-exponential distribution ($\eta=0$) with a uniform background noise. The set of unknown parameters is $\bm{\theta} = (N,\Gamma)$. \Fig{fig1} shows the dependence of the $F$-value upon the number of counts due to background noise (\fig{fig1}(a)), the number of data points per lifetime (\fig{fig1}(b)), the number of fluorescence lifetimes per repetition period (\fig{fig1}(c)) and the standard deviation of the IRF (\fig{fig1}(d)) which is assumed to follow a inverse Gaussian distribution. All these parameters fully characterize the experimental conditions, and must be optimized in order to perform shot-noise limited estimations.

An analytical expression of $F$ was obtained by K\"ollner and Wolfrum~\cite{kollner_how_1992}, for the special case in which the IRF is modelled by a Dirac delta function and the number of counts due to background noise is $\beta=0$:
\begin{equation}
F(r,k) = \frac{k}{r} \, \sqrt{1-e^{-r}} \left[ \frac{e^{r/k} (1-e^{-r})}{(e^{r/k}-1)^2}- \frac{k^2}{e^r-1} \right]^{-1/2} \; .
\label{eq11}
\end{equation}
As expected, the results obtained from \eq{eq11} and those obtained by numerically inverting the information matrix are the same under these conditions, as shown in \fig{fig1}(b) and \fig{fig1}(c).

The results shown in \fig{fig1} can be straightforwardly applied to identify the parameters limiting the precision of an experimental setup. For instance, if $N=400$ fluorescence photons are detected from an emitter, it follows from \eq{eq9} that a relative error on $\Gamma$ of 6\% can be reached for $F\leq 1.2$. Considering each of the four situations depicted in \fig{fig1} individually, such precision can be achieved for a number of counts due to the background noise $\beta\leq10^{-2}$ (\fig{fig1}(a)), for a number of data points per lifetime $k\geq0.5$ (\fig{fig1}(b)), for a number of fluorescence lifetimes per repetition period $r\geq4$ (\fig{fig1}(c)), and for a standard deviation of the IRF $\Gamma \sigma_{irf}\leq0.7$ (\fig{fig1}(d)). When these four conditions on the parameters are simultaneously verified, one is ensured to obtain a lower bound on $\sigma_{\Gamma_1}/\Gamma_1$ smaller than 9.7\%.

\bigskip

\subsection{Bi-exponential case}

\begin{figure*}[htbp]
\begin{center}
\includegraphics[scale=0.74]{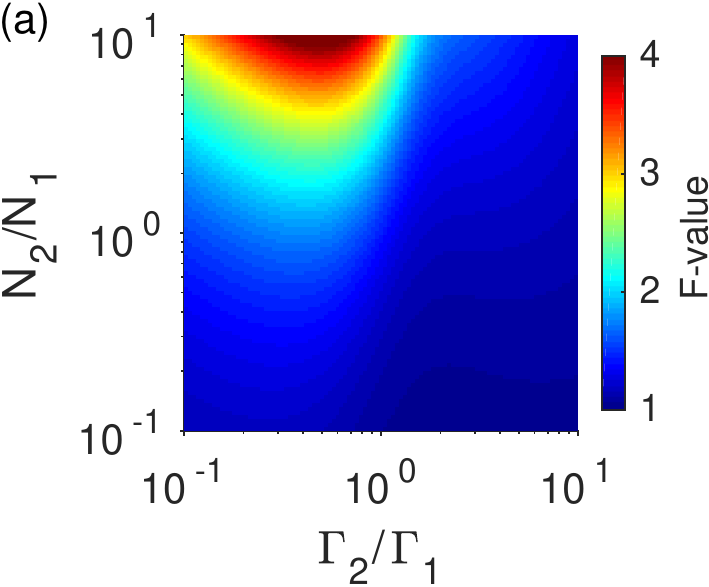} \hspace{0.4cm}
\includegraphics[scale=0.74]{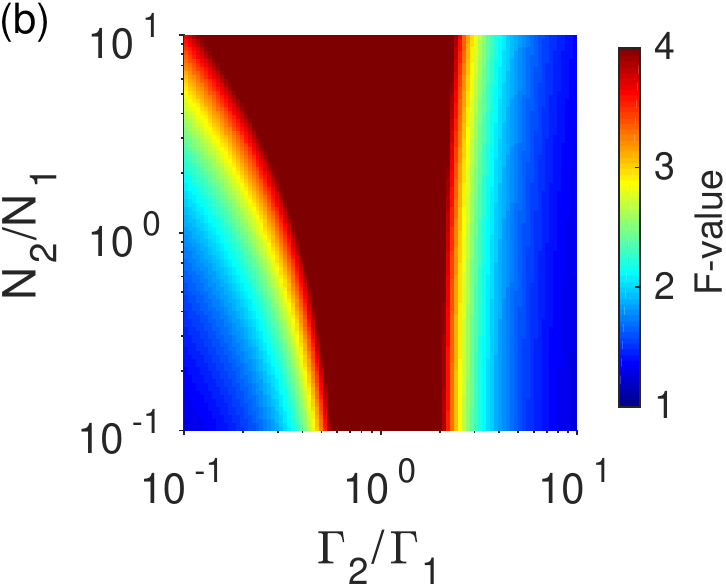}\hspace{0.4cm}
\includegraphics[scale=0.74]{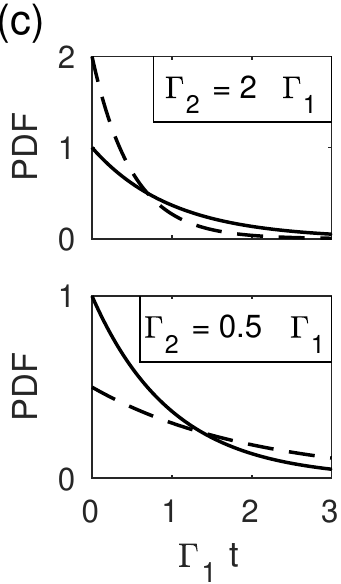}
\end{center}
\caption{Precision of lifetime estimations as a function of $\eta=N_2/N_1$ and $\gamma=\Gamma_2/\Gamma_1$ for a bi-exponential distribution when (a) the parameters $N_2$ and $\Gamma_2$ are known and when (b) these parameters must be estimated from the data. (c) Probability density function associated with the first component (solid lines) and with the second component (dashed lines) for $\Gamma_2 = 2 \, \Gamma_1$ and $\Gamma_2 = 0.5 \, \Gamma_1$. 
}
\label{fig2}
\end{figure*}

Let us now consider a signal following a bi-exponential distribution. If $N_2$ and $\Gamma_2$ can be precisely estimated from independent measurements, the set of unknown parameters is $\bm{\theta} = (N_1,\Gamma_1)$. Such situation can occur for instance if the second component originates from the luminescence of the substrate, because $N_2$ and $\Gamma_2$ can be estimated by performing a reference measurement without the emitter. In contrast, if the second component originates from the emitter itself, $N_2$ and $\Gamma_2$ are generally estimated from the data, and the set of unknown parameters to be considered becomes $\bm{\theta} = (N_1,\Gamma_1,N_2,\Gamma_2)$.

In both cases, the precision of lifetime estimations depends on $\eta=N_2/N_1$ and $\gamma=\Gamma_2/\Gamma_1$. \Fig{fig2} shows the dependence of the $F$-value upon these parameters for both situations. As we aim at estimating $\Gamma_1$, the second decay should be interpreted as a noise source which can only degrade the estimation precision. Hence, the estimation precision is best for $N_2<N_1$. Interestingly, when the parameters of the second component can be independently estimated (\fig{fig2}(a)), the estimation precision is best when $\Gamma_2>\Gamma_1$, \textit{i.e.} the lifetime of the second component is shorter than the lifetime to be estimated. Indeed, $\Gamma_1$ can be correctly estimated from the last points of the decay histogram, for which the contribution of the second component has vanished (see Section~3 of \textit{Appendix: Numerical methods}). This arguments also hold when the parameters of the second component must be estimated from the data (\fig{fig2}(b)) but, in that case, $\Gamma_1$ and $\Gamma_2$ must be significantly different in order to enable a correct estimation.

As an example, we can investigate the number of fluorescence photons required to obtain a relative error of 6\% if the number of photons detected from each component is the same ($N_2/N_1=1$) and for the two situations represented in \fig{fig2}(c). To begin with, we consider that the parameters of the second component are independently estimated. While $N_1=381$ photons are required if $\Gamma_2=2 \,\Gamma_1$, $N_1=852$ photons are needed if $\Gamma_2=0.5 \, \Gamma_1$. This example illustrates that the situation $\Gamma_2>\Gamma_1$ is favorable in order to precisely estimate the decay rate~$\Gamma_1$. As expected, the number of photons required to properly estimate $\Gamma_1$ drastically increases when the parameters of the second component must be estimated from the data. Indeed, $N_1=5,237$ photons are required if $\Gamma_2=2 \,\Gamma_1$ and $N_1=11,450$ photons are required if $\Gamma_2=0.5 \, \Gamma_1$.

\subsection{Numerical tests}

In this section, we demonstrate the versatility of the approach by analyzing two examples, one for the estimation of lifetimes on the order of several picoseconds, the other for the estimation of lifetimes on the order of the nanosecond. In both cases, we consider an exponential decay with known background noise, so that the parameters to be estimated are $\bm{\theta}=(N,\Gamma)$. Each example takes into account an IRF that was experimentally measured on a TCSPC setup. In these two examples, we compare the Cram\'er-Rao bound on lifetime estimators to numerical results obtained from Monte Carlo experiments. To this end, we numerically generate a set of 10,000 decay histograms for each experimental condition that will be investigated. This is performed by using \eq{eq2} to calculate the cumulative distribution function, from which decay histograms can be randomly generated based on the inversion principle \cite{devroye_inversion_1986}. The lifetime $\tau=1/\Gamma$ is then estimated from each histogram by using a maximum-likelihood (ML) method. The estimation bias is not significant for these numerical experiments (see Section~4 of \textit{Appendix: Numerical methods}). Moreover, the variance of ML estimators asymptotically approaches the Cram\'er-Rao bound for large sample statistics, which allows the root-mean square (RMS) deviation of the estimated lifetimes to be close to the fundamental limit for the experimental conditions that are investigated. Note that the implementation of the second example is provided in the open source Code File 1 \cite{code_ref}.

\begin{figure*}[htbp]
\begin{center}
\includegraphics[scale=0.72]{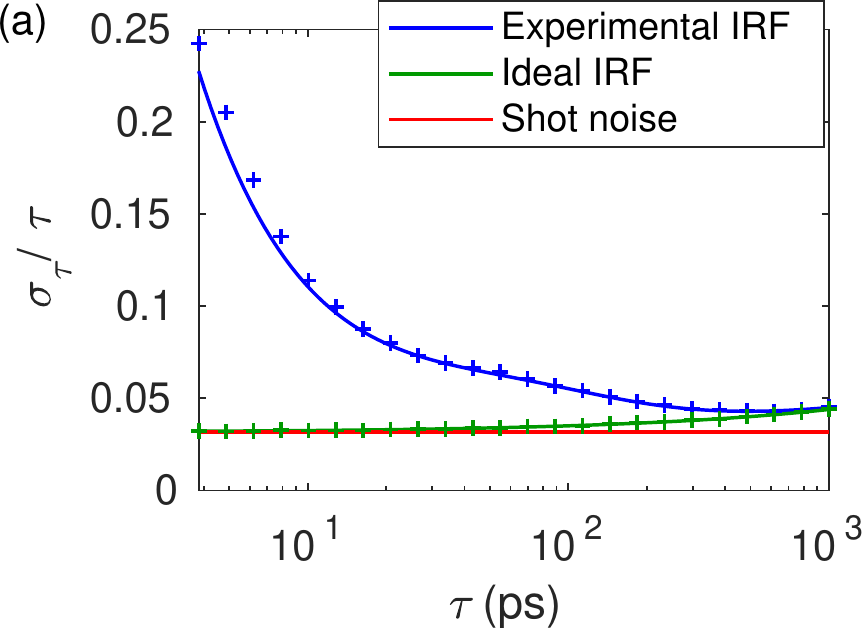} \hspace{0.5cm}
\includegraphics[scale=0.72]{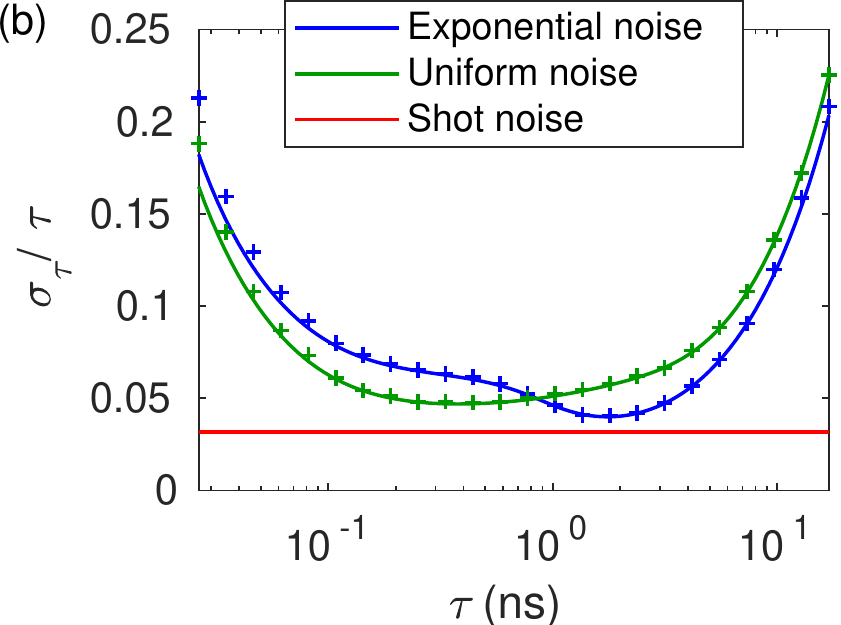}
\end{center}
\caption{Relative standard error on lifetime estimates for an experimental setup designed for (a) picosecond and (b) nanosecond lifetime estimations. Solid lines represent the lower bounds on $\sigma_\tau/\tau$ and data points are the RMS deviations of lifetime estimates obtained from randomly generated data.}
\label{fig3}
\end{figure*}

\paragraph{Precision of picosecond lifetime estimations}

We consider an experimental setup designed to estimate picosecond lifetimes, with a board resolution of 1~ps, with a repetition rate of 80~MHz ($T=12.5$~ns) and with an IRF characterised by a full with at half maximum (FWHM) of 38~ps. The expected number of fluorescence photons is set to $N=1,000$. Moreover, the background signal is assumed to be uniform, and the expected number of detections due to this background is set to $N_b=500$. We calculated the lower bound on the relative standard error $\sigma_\tau/\tau$ as a function of the lifetime for these experimental conditions (\fig{fig3}(a), blue curve) as well as in the case the IRF is a Dirac delta function (\fig{fig3}(a), green curve). For $\tau\sim1$~ns, the lower bound on $\sigma_\tau/\tau$ is slightly larger than the shot-noise limited bound due to the background noise. For an ideal IRF, the lower bound on $\sigma_\tau/\tau$ slightly decreases when $\tau$ decreases since the number of counts per lifetime due to background noise (noted $\beta$) decreases. In contrast, when the actual IRF is considered, the lower bound on $\sigma_\tau/\tau$ strongly increases for lifetime shorter than half the FWHM, which clearly highlights the relevance of taking into account the IRF for picosecond lifetime estimations. Note that, in this regime, the ML estimator becomes less efficient and slightly deviates from the Cram\'er-Rao bound.

\paragraph{Optimal contrast in fluorescence lifetime microscopy}

We consider a typical TCSPC setup designed for nanosecond lifetime estimation, with a board resolution of 16~ps, a repetition rate of 80~MHz and an IRF characterized by a FWHM of 240~ps. The expected number of fluorescence photons is set to $N=1,000$. Moreover, we assume that a luminescence background signal is also detected, with the same intensity as the fluorescence signal of interest and a lifetime of 1~ns. In order to choose fluorescent emitters that will lead to the most contrasted FLIM images, we calculated the lower bound on $\sigma_\tau/\tau$ as a function of the fluorescence lifetime of the emitter under these experimental conditions (\fig{fig3}(b), blue curve) as well as in the case of uniform background noise (\fig{fig3}(b), green curve). While the precision of lifetime estimations is optimal for $\tau\sim400$~ps for the uniform background, the precision of lifetime estimations is optimal for $\tau\sim2$~ns for the exponential background. Indeed, the precision of lifetime estimations depends on $\gamma$ (see \fig{fig2}(a)) and the time dependence of the background noise allows estimations that are more precise for $\gamma>1$, that is, $\tau>1$~ns. In both situations, for lifetimes larger than 5~ns, the lower bound on $\sigma_\tau/\tau$ strongly increases as the number of fluorescence lifetimes per repetition period (noted $r$) decreases. In this regime, using a laser with a lower repetition rate must be considered in order to improve the precision of lifetime estimations. Moreover, for lifetimes smaller than 100~ps, the lower bound on $\sigma_\tau/\tau$ also strongly increases as the fluorescence lifetime becomes comparable to the IRF. In this regime, the precision of the estimations could be improved by using a detection system with a smaller jitter and an excitation laser with a shorter pulse width.

\section{Conclusion}

In summary, we calculated the lower bound on the standard error on lifetime estimates depending on key experimental parameters. These results can be used as a benchmark for the evaluation of the precision of lifetime estimations. Moreover, they reveal the influence of different parameters upon the estimation precision, providing us with a powerful tool for the optimisation of a TCSPC setup as illustrated by two examples. We notably showed that a significant enhancement of the precision of lifetime estimations can be achieved by choosing the proper fluorescent emitter depending on the expected background noise, the IRF of the setup and the repetition rate of the excitation laser. Various dependent parameters such as the integration time, the power of the excitation laser or the measured spectral range can easily be studied by using the proposed formalism, which can also be extended to the study of time-gated photon counting techniques. We expect these results to be of great interest for current experimental challenges such as the reduction of the acquisition time in FLIM-based techniques, the characterization of photonic antennas with high Purcell factors as well as fluorescence lifetime measurements at the single molecule level.

\section{Appendix: Numerical methods}

In this section, we give explicit expressions of the Fisher information matrix and of the Cram\'er-Rao bound, and we provide additional numerical results.

\subsection{Calculation of the information matrix}

\subsubsection{General expression}
\allowdisplaybreaks
Using the dimensionless parameters defined in the manuscript, the expectation of each data item reads

\begin{equation}
\begin{split}
f_i =& N_1 \left( \sum_{l=0}^{+ \infty} \int\limits_{u_i+lr}^{u_{i+1}+lr} \left[ \tilde{q}_{irf}(u) \ast e^{-u} \right] \de u \right. \\
  +& \left. \eta \sum_{l=0}^{+ \infty} \int\limits_{u_i+lr}^{u_{i+1}+lr} \left[ \tilde{q}_{irf}(u) \ast \gamma e^{-\gamma u} \right] \de u \right. \\
 +& \left. \beta r \int\limits_{u_i}^{u_{i+1}} \tilde{q}_{b}(u) \de u \right) \; .
\end{split}
\label{eq_appendix_1}
\end{equation}

In order to calculate the information matrix, let us define $\bm{J}^{\mathrm{I}}$, $\bm{J}{^\mathrm{II}}$, $\bm{K}{^\mathrm{I}}$, $\bm{K}{^\mathrm{II}}$ and $\bm{J}{^\mathrm{B}}$ as follows:
\begin{align}
J^{\mathrm{I}}_i &= \sum_{l=0}^{+ \infty} \int\limits_{u_i+lr}^{u_{i+1}+lr} \left[ \tilde{q}_{irf}(u) \ast e^{- u} \right] \de u \; , \nonumber \\
K^{\mathrm{I}}_i &= \sum_{l=0}^{+ \infty} \int\limits_{u_i+lr}^{u_{i+1}+lr} \left[ \tilde{q}_{irf}(u) \ast (1-u) e^{- u} \right] \de u \; , \nonumber \\
J^{\mathrm{II}}_i &= \sum_{l=0}^{+ \infty} \int\limits_{u_i+lr}^{u_{i+1}+lr} \left[ \tilde{q}_{irf}(u) \ast \gamma e^{- \gamma u} \right] \de u \; , \label{eq_appendix_2} \\
K^{\mathrm{II}}_i &= \sum_{l=0}^{+ \infty} \int\limits_{u_i+lr}^{u_{i+1}+lr} \left[ \tilde{q}_{irf}(u) \ast \gamma (1-\gamma u) e^{- \gamma u} \right] \de u \; , \nonumber \\
J^{\mathrm{B}}_i &= r \int\limits_{u_i}^{u_{i+1}} \tilde{q}_{b}(u) \de u \; . \nonumber
\end{align}
With these notations, \eq{eq_appendix_1} reads $f_i = N_1 \left( J^{\mathrm{I}}_i + \eta J^{\mathrm{II}}_i + \beta J^{\mathrm{B}}_i \right)$. Differentiating this expression by each parameters yields
\begin{align}
\frac{\partial f_i}{\partial N_1}& = J^{\mathrm{I}}_i \; , \nonumber \\
\frac{\partial f_i}{\partial \Gamma_1}&= \frac{N_1}{\Gamma_1} K^{\mathrm{I}}_i \; , \nonumber \\
\frac{\partial f_i}{\partial N_b}&= J^{\mathrm{B}}_i \; ,  \\
\frac{\partial f_i}{\partial N_2}&= J^{\mathrm{II}}_i \; ,\nonumber \\
\frac{\partial f_i}{\partial \Gamma_2}&= \frac{\eta N_1}{\gamma \Gamma_1} K^{\mathrm{II}}_i \; \nonumber .
\end{align}
Let us recall the expression of the information matrix:
\begin{equation}
\left[\bm{\mathcal{I}}(\bm{\theta})\right]_{jk} = \sum_{i=1}^n \frac{1}{f_i}\left( \frac{\partial f_i}{\partial \theta_j} \right) \left( \frac{\partial f_i}{\partial \theta_k} \right) \; .
\label{eq_appendix_3}
\end{equation}
where $\bm{\theta}=(N_1,\Gamma_1,N_b,N_2,\Gamma_2)$ are the parameters to be estimated. The elements of the information matrix are therefore expressed by
\begin{align}
\mathcal{I}_{N_1N_1} & = \frac{1}{N_1} \; \sum_{i=1}^n \frac{(J^{\mathrm{I}}_i )^2}{J^{\mathrm{I}}_i + \eta J^{\mathrm{II}}_i+ \beta J^{\mathrm{B}}_i} \; , \nonumber \\
\mathcal{I}_{\Gamma_1 \Gamma_1} & = \frac{N_1 }{\Gamma_1^2} \; \sum_{i=1}^n \frac{(K^{\mathrm{I}}_i )^2}{J^{\mathrm{I}}_i + \eta J^{\mathrm{II}}_i+ \beta J^{\mathrm{B}}_i} \; , \nonumber \\
\mathcal{I}_{N_1 \Gamma_1} & = \frac{1}{\Gamma_1} \; \sum_{i=1}^n \frac{J^{\mathrm{I}}_i K^{\mathrm{I}}_i}{J^{\mathrm{I}}_i + \eta J^{\mathrm{II}}_i+ \beta J^{\mathrm{B}}_i} \; , \nonumber \\
\mathcal{I}_{N_b N_b} & = \frac{1}{N_1} \; \sum_{i=1}^n \frac{ (J^{\mathrm{B}}_i)^2}{J^{\mathrm{I}}_i + \eta J^{\mathrm{II}}_i+ \beta J^{\mathrm{B}}_i} \; , \nonumber \\
\mathcal{I}_{N_1 N_b} & = \frac{1}{N_1} \; \sum_{i=1}^n \frac{ J^{\mathrm{I}}_i J^{\mathrm{B}}_i}{J^{\mathrm{I}}_i + \eta J^{\mathrm{II}}_i+ \beta J^{\mathrm{B}}_i} \; , \nonumber \\
\mathcal{I}_{\Gamma_1 N_b} & = \frac{1}{\Gamma_1} \; \sum_{i=1}^n \frac{J^{\mathrm{B}}_i K^{\mathrm{I}}_i }{J^{\mathrm{I}}_i + \eta J^{\mathrm{II}}_i+ \beta J^{\mathrm{B}}_i} \; , \nonumber \\
\mathcal{I}_{N_2 N_2} & = \frac{1}{N_1} \; \sum_{i=1}^n \frac{(J^{\mathrm{II}}_i )^2}{J^{\mathrm{I}}_i + \eta J^{\mathrm{II}}_i+ \beta J^{\mathrm{B}}_i} \; , \nonumber \\
\mathcal{I}_{N_1N_2} & = \frac{1}{N_1} \; \sum_{i=1}^n \frac{J^{\mathrm{I}}_i J^{\mathrm{II}}_i }{J^{\mathrm{I}}_i + \eta J^{\mathrm{II}}_i+ \beta J^{\mathrm{B}}_i} \; , \\
\mathcal{I}_{\Gamma_1N_2} & = \frac{1}{\Gamma_1} \; \sum_{i=1}^n \frac{J^{\mathrm{II}}_i K^{\mathrm{I}}_i }{J^{\mathrm{I}}_i + \eta J^{\mathrm{II}}_i+ \beta J^{\mathrm{B}}_i} \; , \nonumber \\
\mathcal{I}_{N_b N_2} & = \frac{1}{N_1} \;\sum_{i=1}^n \frac{ J^{\mathrm{B}}_i J^{\mathrm{II}}_i }{J^{\mathrm{I}}_i + \eta J^{\mathrm{II}}_i+ \beta J^{\mathrm{B}}_i} \; , \nonumber \\
\mathcal{I}_{\Gamma_2 \Gamma_2} & = \frac{\eta^2 N_1}{\gamma^2 \Gamma_1^2} \; \sum_{i=1}^n \frac{(K^{\mathrm{II}}_i )^2}{J^{\mathrm{I}}_i + \eta J^{\mathrm{II}}_i+ \beta J^{\mathrm{B}}_i} \; , \nonumber \\
\mathcal{I}_{N_1 \Gamma_2} & = \frac{\eta}{\gamma \Gamma_1} \sum_{i=1}^n \frac{J^{\mathrm{I}}_i K^{\mathrm{II}}_i }{J^{\mathrm{I}}_i + \eta J^{\mathrm{II}}_i+ \beta J^{\mathrm{B}}_i} \; , \nonumber \\
\mathcal{I}_{\Gamma_1\Gamma_2} & = \frac{\eta N_1}{\gamma \Gamma_1^2} \; \sum_{i=1}^n \frac{ K^{\mathrm{I}}_i K^{\mathrm{II}}_i }{J^{\mathrm{I}}_i + \eta J^{\mathrm{II}}_i+ \beta J^{\mathrm{B}}_i} \; , \nonumber \\
\mathcal{I}_{N_b \Gamma_2} & = \frac{\eta}{\gamma \Gamma_1} \; \sum_{i=1}^n \frac{ J^{\mathrm{B}}_i K^{\mathrm{II}}_i }{J^{\mathrm{I}}_i + \eta J^{\mathrm{II}}_i+ \beta J^{\mathrm{B}}_i} \; , \nonumber \\
\mathcal{I}_{N_2 \Gamma_2} & = \frac{\eta}{\gamma \Gamma_1} \; \sum_{i=1}^n \frac{J^{\mathrm{II}}_i K^{\mathrm{II}}_i }{J^{\mathrm{I}}_i + \eta J^{\mathrm{II}}_i+ \beta J^{\mathrm{B}}_i} \; . \nonumber
\end{align}
Since $\Gamma_1$ is our reference, the Cram\'er-Rao inequality will be most conveniently expressed in terms of this parameter. By inverting the information matrix, we indeed obtain the following expression:
\begin{equation}
\frac{\sigma_{\Gamma_1}}{\Gamma_1} \geq \frac{1}{\sqrt{N_1}} \times F \left(\eta,\gamma,r,k,\tilde{q}_{irf},\beta,\tilde{q}_{b} \right) \; ,
\label{eq_appendix_9}
\end{equation}
where $\sigma_{\Gamma_1}$ is the standard error on the decay rate estimates and $F$ can be calculated by numerical inversion of the information matrix.

\subsubsection{Limiting cases}

\paragraph{Ideal IRF}

Whenever the IRF can be considered as a Dirac delta function, the coefficients defined by \eq{eq_appendix_2} become
\begin{equation}
\begin{split}
J^{\mathrm{I}}_i =& \left[ e^{-u_{i}} - e^{-u_{i+1}} \right] \sum_{l=0}^{+ \infty} e^{-lr} \; , \\
K^{\mathrm{I}}_i =&\left[u_{i+1} e^{-u_{i+1}} -u_{i} e^{-u_{i}} \right] \sum_{l=0}^{+ \infty} e^{-lr} \\
&+ \left[ e^{-u_{i+1}} -e^{-u_{i}} \right] \sum_{l=0}^{+ \infty} lr e^{-lr} \; , \\
J^{\mathrm{II}}_i =& \left[ e^{-\gamma u_i}-e^{- \gamma u_{i+1}} \right] \sum_{l=0}^{+ \infty} e^{-\gamma lr} \; , \\
K^{\mathrm{II}}_i =&\left[\gamma u_{i+1} e^{-\gamma u_{i+1}} - \gamma u_{i} e^{-\gamma u_{i}} \right] \sum_{l=0}^{+ \infty} e^{-\gamma lr} \\
&+ \left[ e^{-\gamma u_{i+1}} -e^{-\gamma u_{i}} \right] \sum_{l=0}^{+ \infty} \gamma lr e^{-\gamma lr} \; . 
\end{split}
\end{equation}
$J^{\mathrm{B}}_i$ remains unchanged, as it does not depend on the IRF. These expressions can be further simplified, by using the following properties of geometric series:
\begin{equation}
\begin{split}
 \sum_{l=0}^{+ \infty} e^{-lr} &= \frac{1}{1-e^{-r}} \; , \\
 \sum_{l=0}^{+ \infty} l r e^{-lr} &= \frac{r e^{-r}}{(1-e^{-r})^2} \; . \\
\end{split}
\end{equation}
We obtain
\begin{equation}
\begin{split}
J^{\mathrm{I}}_i =& \frac{e^{-u_{i}} - e^{-u_{i+1}}}{1-e^{-r}} \; , \\
K^{\mathrm{I}}_i =& \frac{u_{i+1} e^{-u_{i+1}} -u_{i} e^{-u_{i}}}{1-e^{-r}} \\
&+ \frac{r e^{-r} \left( e^{-u_{i+1}} -e^{-u_{i}} \right)}{(1-e^{-r})^2} \; , \\
J^{\mathrm{II}}_i =& \frac{e^{-\gamma u_i}-e^{- \gamma u_{i+1}}}{1-e^{-\gamma r}} \; , \\
K^{\mathrm{II}}_i =& \frac{\gamma u_{i+1} e^{-\gamma u_{i+1}} - \gamma u_{i} e^{-\gamma u_{i}}}{1-e^{-\gamma r}} \\
&+ \frac{\gamma r e^{-\gamma r} \left( e^{-\gamma u_{i+1}} -e^{-\gamma u_{i}} \right) }{(1-e^{-\gamma r})^2} \; .
\end{split}
\end{equation}

\paragraph{Uniform background noise}

Whenever the background noise is uniform over the repetition period, the coefficient $J^{\mathrm{B}}_i$ simply becomes
\begin{equation}
J^{\mathrm{B}} = r /n \; .
\end{equation}

\subsection{Cram\'er-Rao bound on the lifetime estimator}

From \eq{eq_appendix_9}, it is straightforward to calculate the Cram\'er-Rao bound on the lifetime estimator. To do so, we define the excited-state lifetime $\tau_1=1/\Gamma_1$ and we perform a transformation of parameter, as detailed in Appendix~3B of \cite{kay_cramer-rao_1993}. This reads 
\begin{equation}
\Ve \left( \hat{\tau}_1 \right) \geq \left[ \frac{\partial (1/\Gamma_1)}{\partial \Gamma_1} \right]^2 \left[ \frac{\Gamma_1}{\sqrt{N_1}} \,  F \left(\eta,\gamma,r,k,\tilde{q}_{irf},\beta,\tilde{q}_{b} \right) \right]^2 \; ,
\end{equation}
where $\hat{\tau}_1$ is the lifetime estimator. This expression simplifies to
\begin{equation}
\frac{\sigma_{\tau_1}}{\tau_1} \geq \frac{1}{\sqrt{N_1}} \times  F \left(\eta,\gamma,r,k,\tilde{q}_{irf},\beta,\tilde{q}_{b} \right) \; ,
\end{equation}
where $\sigma_{\tau_1}$ is the standard error on the lifetime estimates. This demonstrates that the Cram\'er-Rao lower bounds are the same for the relative standard error on the decay rate and lifetime estimators.

\subsection{Partial and cumulated information}

In general, elements of the information matrix are calculated as a sum over the $n$ points of the decay histogram. From \eq{eq_appendix_3}, it can therefore be expressed in a formal way as follows:
\begin{equation}
\bm{\mathcal{I}} =  \sum_{i=1}^n \bm{\mathcal{R}}_i(\eta, \gamma, r,k,\tilde{q}_{irf},\beta,\tilde{q}_{b})\; .
\label{eq_appendix_10}
\end{equation}
We define the partial information matrix $\bm{\mathcal{I}}_p$ as the information matrix calculated from the sum over $p$ points of the decay histogram, where $p$ is the number of parameters to be estimated. This reads
\begin{equation}
\bm{\mathcal{I}}_p =  \sum_{i=j}^{j+(p-1)} \bm{\mathcal{R}}_i(\eta, \gamma, r,k,\tilde{q}_{irf},\beta,\tilde{q}_{b})\; .
\end{equation}
We also define the cumulated information matrix $\bm{\mathcal{I}}_c$ as the information matrix calculated from the sum over the $m$ first points of the decay histogram, with $m \leq n$. This reads
\begin{equation}
\bm{\mathcal{I}}_c =  \sum_{i=1}^{m} \bm{\mathcal{R}}_i(\eta, \gamma, r,k,\tilde{q}_{irf},\beta,\tilde{q}_{b})\; .
\end{equation}

\bigskip
\begin{figure*}[ht]
\begin{center}
\includegraphics[scale=0.77]{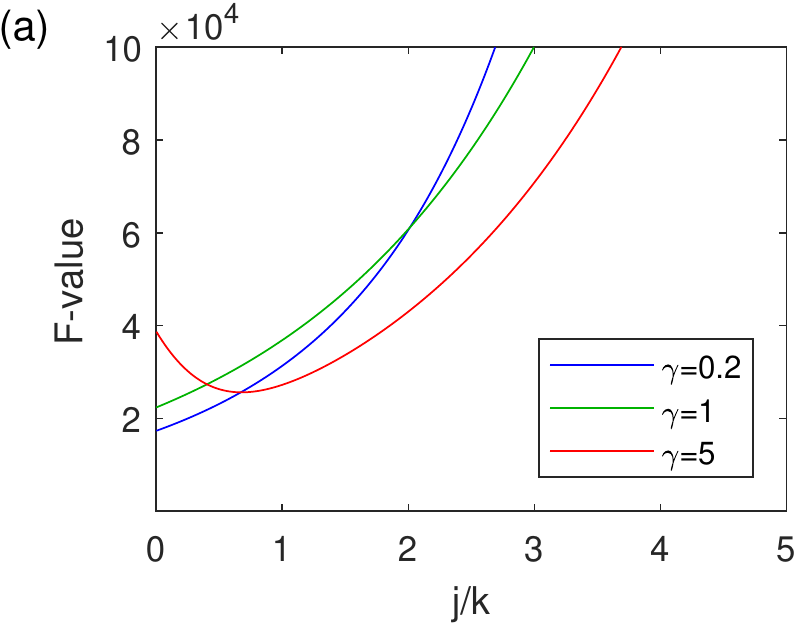} \hspace{0.5cm}
\includegraphics[scale=0.77]{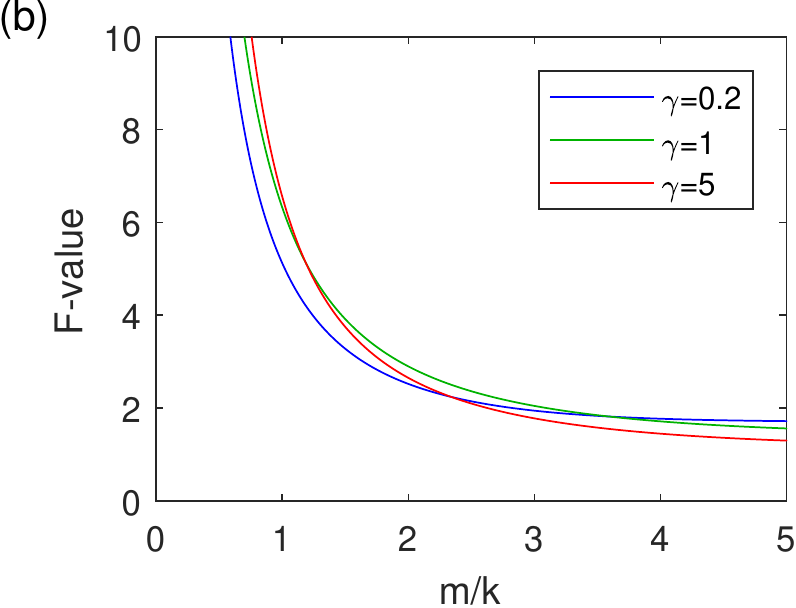}
\end{center}
\caption{Precision of lifetime estimations calculated for a bi-exponential distribution assuming that the parameters of the second decay component are known. (a)~$F$-value calculated from the partial information matrix, as a function of $j/k$. (b)~$F$-value calculated from the cumulated information matrix, as a function of $m/k$. 
}
\label{fig_app_1}
\end{figure*}
\bigskip

From these definitions, the partial information and the cumulated information can be used to compare the information carried by different sets of points of the data histogram about the decay rate $\Gamma_1$ to be estimated. \Fig{fig_app_1}(a) shows the $F$-value calculated from the partial information matrix as a function of $j/k$, considering a bi-exponential decay histogram ($\eta=1$) with a large number of data point per lifetime ($k=500$), a large number of lifetime per repetition period ($r=100$), no background noise ($\beta=0$) and an ideal IRF. The first points of the decay histogram carry more information for $\gamma=0.2$ than for $\gamma=5$. Indeed, in this latter case, the signal due to the second decay is concentrated on the first points of the decay histogram, making the estimation of $\Gamma_1$ more difficult. However, this signal vanishes faster and, indeed, we can see that the information contained in the last points of the histogram is larger for $\gamma=5$ than for $\gamma=0.2$. \Fig{fig_app_1}(b) shows the $F$-value calculated from the cumulated information matrix as a function of $m/k$. For $m/k \sim 2.3$, the $F$-value becomes smaller for $\gamma=5$ than for $\gamma=0.2$, resulting in a better precision for $\gamma=5$ than for $\gamma=0.2$ whenever the whole histogram is considered.

\subsection{Bias of maximum-likelihood estimations}

\begin{figure}[ht]
\begin{center}
\includegraphics[scale=0.89]{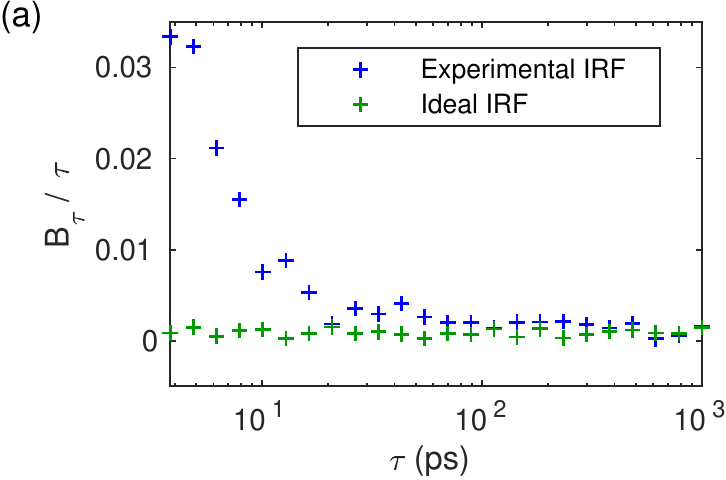}\\
\includegraphics[scale=0.89]{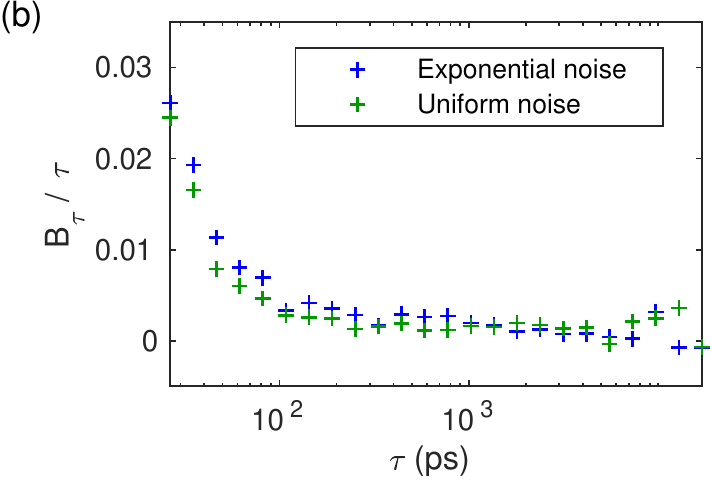}
\end{center}
\caption{Relative bias on lifetime estimates for an experimental setup designed for (a) picosecond and (b) nanosecond lifetime estimations.}
\label{fig_app_2}
\end{figure}

The bias $B_\tau$ of an estimator $\hat{\tau}$ describes whether the estimator can, on average, recover the true value of the parameter $\tau$. It is defined by 
\begin{equation}
 B_\tau=\Ee (\hat{\tau})-\tau \; .
\end{equation}
\Fig{fig_app_2} shows the relative bias $B_\tau/\tau$ of the maximum-likelihood estimator used in the two examples analyzed in the manuscript. In both cases, we can see that the relative bias increases when $\tau$ decreases, which can be attributed to the influence of the IRF on the estimation process. However, the relative bias is always one order of magnitude smaller than the relative standard error (Fig.~3 of the manuscript). This justifies the relevance of using the Cram\'er-Rao bound as a benchmark for the estimation precision (the Cram\'er-Rao bound applies only to unbiased estimators).

\bigskip
\bigskip

\section*{Funding}
Ville de Paris (Emergences 2015); Agence Nationale de la Recherche (ANR) (ANR-10\_LABX-24, ANR-10- IDEX-0001-02 PSL*, ANR-17-CE09-0006).

\section*{Acknowledgments}

The authors thank Hans Gerritsen for helpful comments about the manuscript and Sebastien Bidault for providing the IRF used in the first example.

\section*{Disclosures}
The authors declare that there are no conflicts of interest related to this article.

%\bibliographystyle{apsrev_no_url}
%\bibliographystyle{apsrev}
%\bibliography{references}

\end{document}